\documentclass[aps,prl,preprint,final,letterpaper]{revtex4}

\usepackage{graphicx}   
\usepackage{import}                         
\usepackage{epstopdf}
\usepackage{amsmath} 
\usepackage{bm}
\usepackage{amssymb}
\usepackage{quotes}
\usepackage{color}
\usepackage{transparent}
\usepackage{dcolumn}
\usepackage{multirow}
\usepackage{cancel} 
\usepackage{mdframed}
\usepackage{color}
\usepackage{bm}
\usepackage{dsfont}
\usepackage{soul, color} 
\soulregister\ref{7}  
\soulregister\cite{7} 
\renewcommand{\st}[1]{}

\begin{document}
\rmfamily

\title{A near-unity efficiency source of entangled guided waves}
\author{Nicholas Rivera, Ido Kaminer, \& Marin Solja\v{c}i\'{c}}

\affiliation{Department of Physics, MIT, Cambridge, MA 02139, USA \\
$\dagger$ Corresponding author e-mail: nrivera@mit.edu}
\clearpage

\newpage

\begin{abstract}
  Surface phonon polaritons are hybrid modes of photons and optical phonons that can propagate on the surface of a polar dielectric. In this work, we show that the precise combination of confinement and bandwidth offered by surface phonon polaritons allows for the ability to create highly efficient sources of entangled light in the IR/THz. Specifically, phonon polaritons can cause emitters to preferentially decay by the emission of pairs of surface phonon polaritons, instead of the previously dominant single-photon emission. We show that such two-photon emission processes can occur on nanosecond time-scales and can be nearly two orders of magnitude faster than competing single-photon transitions, as opposed to being as much as eight to ten orders of magnitude slower in free space. Our results suggest a fundamentally new design strategy for quantum light sources in the IR/THz: ones that prefer to emit a relatively broad spectrum of entangled photons, potentially allowing for new sources of both single and multiple photons.
\end{abstract}

\noindent	

\maketitle

\noindent

A fundamental rule in light-matter interaction is that when an excited electron in an atom has a choice between emitting one photon and emitting two photons simultaneously, it will nearly always decay via the emission of a single-photon \cite{craig1984molecular,berestetskii1982quantum,cohen1992atom}. The reason for this is impedance mismatch, or equivalently, the mismatch in size between an emitter and its emitted radiation. When one applies this idea to two-photon emission processes, one realizes that two-photon emission suffers much more from impedance mismatch than one-photon emission, leading to its relative suppression.

More quantitatively, the radiation resistance or impedance of a dipole radiator is proportional to $\sqrt{\frac{\mu_0}{\epsilon_0}}(a/\lambda)^2$, which up to other fundamental constants is proportional to  $\alpha(a/\lambda)^2$ \cite{eggleston2015optical}. It turns out that the radiation rate of an atomic dipole is precisely proportional to $\alpha(a/\lambda)^2$, where $a$ is the atomic size, $\lambda$ is the wavelength of the emitted light, and $\alpha \approx 1/137$ is the fine-structure constant. In contrast, the rate of a two-photon process scales as $\alpha^2(a/\lambda)^4$ \cite{breit1940metastability,shapiro1959metastability,goppert1929wahrscheinlichkeit} and suffers much more than the one-photon process when impedance is mismatched. In atomic systems, $(a/\lambda) \sim 1/1000$ and thus, two-photon emission in atoms is consistently slower than one-photon emission by more than eight orders of magnitude. It is because of this simple scaling argument that two-photon processes are considered insignificant and can thus almost always be ignored for the purposes of analyzing the dynamics of excited emitters. It is also because of this simple scaling argument that while conventional (one-photon) spontaneous emission engineering is a paradigm in quantum nano-photonics \cite{tame2013quantum,kumar2015tunable,hoang2015ultrafast,andersen2011strongly,tielrooij2015electrical}, similar engineering has not been nearly as actively pursued for two-photon spontaneous emission processes \cite{hayat2008observation,nevet2010plasmonic,hayat2011applications,ota2011spontaneous,munoz2014emitters}. 

Nevertheless, two-photon spontaneous emission processes have several distinctive features which make them very desirable to access. For example, the photons emitted in such a process are entangled due to energy and angular momentum conservation. For this reason, sources of entangled photons (such as spontaneous parametric down-conversion sources) have become a staple in quantum information protocols. From a more fundamental perspective,  the frequency spectrum of two-photon spontaneous emission can be broad, with a spectral width of the order of the transition frequency itself. This is in sharp contrast to one-photon emission, which is spectrally very sharp in the absence of external broadening mechanisms. That implies that if two-photon processes were sufficiently fast, an emitter with discrete energy levels could be a source of light at continuous frequencies, which is in contrast to one of the first things that one learns about quantum mechanics. 

In this work, we propose a new design strategy for fast production of entangled electromagnetic quanta with quantum efficiencies in excess of 90$\%$. In other words, we propose a scheme in which an excited emitter \textit{prefers} to decay via the simultaneous emission of two quanta, \textit{despite the possibility of allowed single-photon decay pathways}. As a special case, we show how phonon-polaritons in hexagonal boron nitride and other polar dielectrics may allow for the design of a new kind of quantum optics source which emits entangled pairs at rates over an order of magnitude larger than any competing one-photon transitions, corresponding to lifetimes approaching 1 ns in atomic-scale two-photon emitters.

\begin{figure}[hh]
\includegraphics[width=160mm]{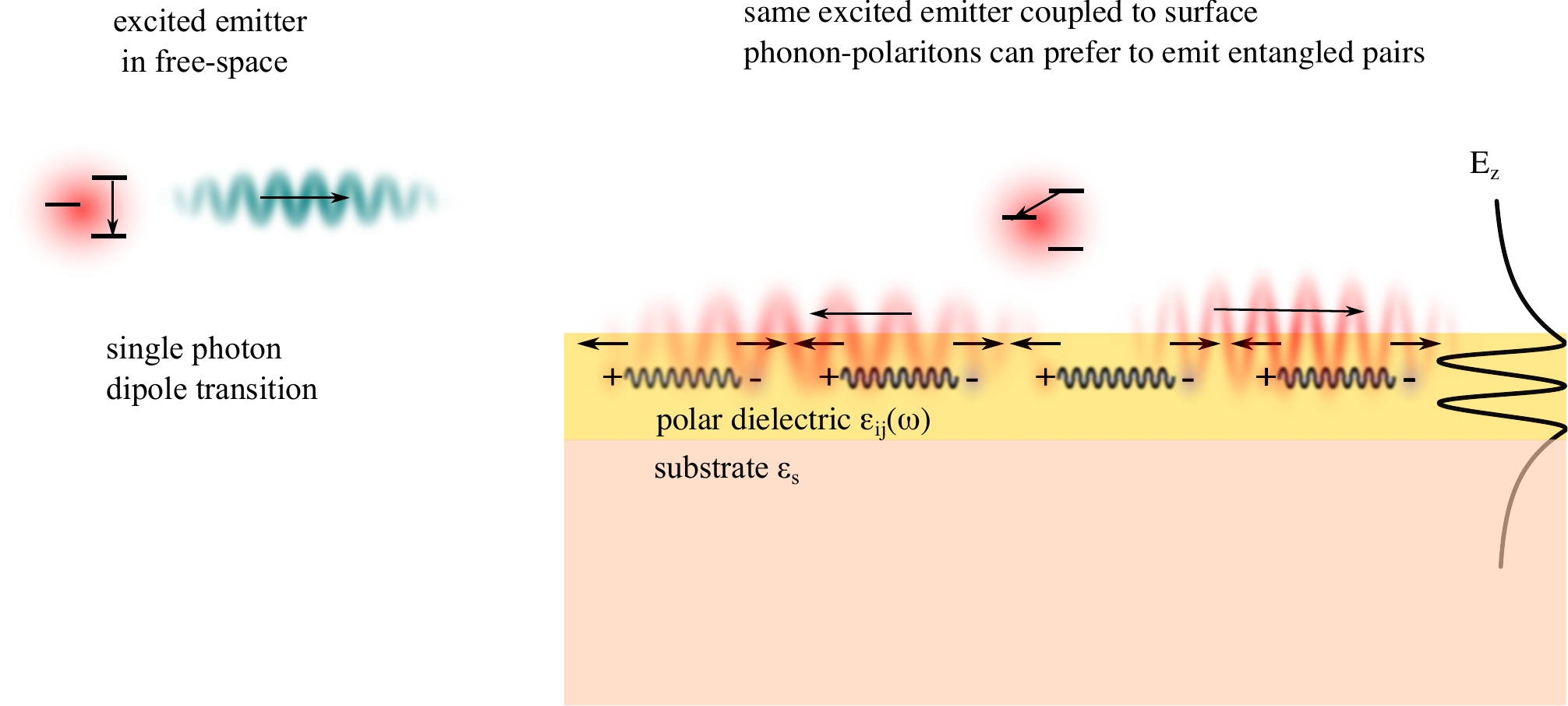}
\caption{\textbf{General Scheme for Accessing Two-Photon Transitions at High-Efficiency:} (Left) Typical situation for an emitter: an emitter may have many choices for a transition, but the relatively high-frequency single-photon dipole transition is chosen. (Right) When coupling that same excited electron to SPhPs in a polar dielectric, the electron can be made to prefer a forbidden transition in the IR (e.g; two-polariton spontaneous emission).}
\end{figure}

The general scheme for accessing high-efficiency forbidden transitions is illustrated in Figure 1. In free-space (and near most nanophotonic structures), an emitter given a choice between different transition pathways will generally take the single-photon E1 transition over other choices. However, if we have highly confined modes (to impedance match) over a sufficiently narrow frequency band, then we can create a situation in which the single-photon E1 transition is negligibly enhanced while the two-photon transition is highly enhanced (over 10 orders of magnitude relative to the enhancement of the competing E1 transition), so much so that the two-photon transition is strongly preferred. As we will now demonstrate, one of the recent material advances that admits the construction of such an emitter is the discovery of highly confined phonon polaritons in polar dielectrics. 

Polar dielectrics like hexagonal boron nitride (hBN), silicon carbide (SiC), lithium niobate (LiNbO$_3$), and others have been the subject of significant attention over the past two years due to their ability to confine electromagnetic energy in small volumes (on length scales potentially as short as 5 nm)~\cite{hillenbrand2002phonon,dai2014tunable,xu2014mid,dai2015graphene,dai2015subdiffractional,li2015hyperbolic,li2016reversible,feng2015localized,tomadin2015accessing,yoxall2015direct,caldwell2015van,caldwell2014sub,caldwell2013low,caldwell2015low}. Moreover, the surface phonon-polaritons (SPhPs) of these materials have extremely strong confinement only over a narrow spectral band of a few THz.  Combined with the fact that these phonon polaritons can have substantially lower losses than other surface excitations like  plasmons \cite{khurgin2015deal}, they may provide a potentially exciting new platform for flatland optics in the technologically interesting IR/THz regime. Much effort has been devoted to demonstrating the coupling and propagation of these modes. In this work, we focus on the potential of using the quantum fluctuations of these modes to match impedance and therefore design quantum emitters in this frequency range which overcome some of the most fundamental limitations of light-matter interactions.

In order to translate our intuition to a quantitative theory, we develop a formalism to compute the rates of two-photon transitions for emitters placed near films supporting SPhP modes (see Methods and Supplementary Materials). Our main assumption is that these phonon-polariton supporting materials are well-described by a temporally local model (though we allow for the permittivitty to be complex and anisotropic). Agreement with a local model has been seen in hBN films as thin as 1 nm, although only the long-wavelength part of the dispersion was measured \cite{dai2014tunable}. Nevertheless, we will see that the results we arrive at should be achievable for thicknesses between 5 and 10 nm, where a local model should certainly suffice.

\section{Results}
\begin{figure}[hh]
\includegraphics[width=160mm]{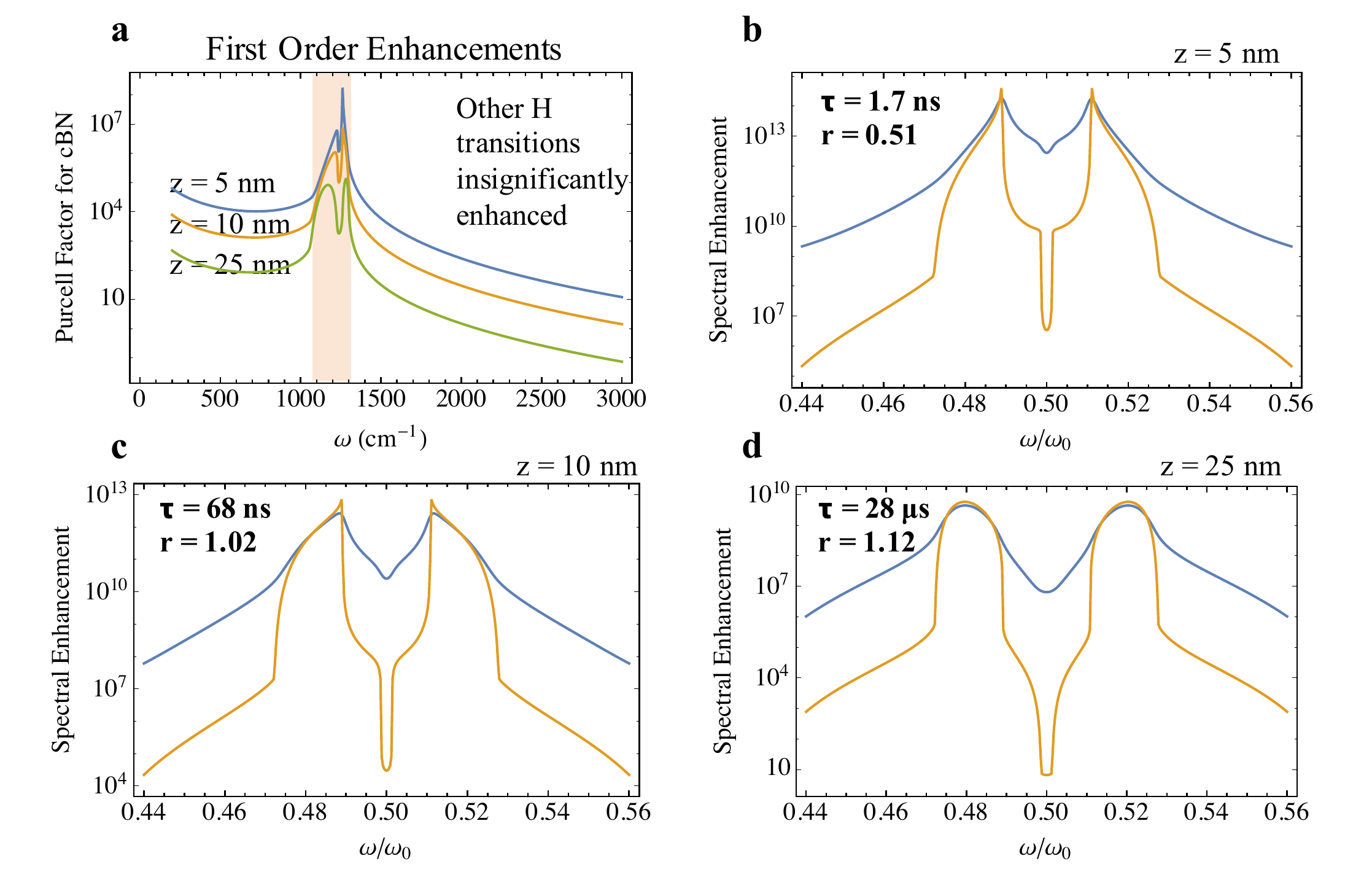}
\caption{\textbf{Making Two-Photon Emission Dominant:} (a) Purcell spectra for a z-polarized dipole above 10 nm thick cBN at atom-surface separations of 5, 10, and 25 nm. (b, c, d). Two-photon Purcell spectra for an s $\rightarrow$ s transition as a function of photon frequency $\omega$ for the same set of atom-surface separations. Blue denotes the Purcell spectra with losses accounted for and the orange denotes the Purcell spectra for cBN with 100x weaker losses. In each of (b,c,d) we note both the overall two-photon transition rate between the 5s and 4s states of hydrogen and the corresponding radiative ratio ($\Gamma_R/\Gamma$). The permittivity of the substrate is taken to be 2 and the damping constant for cBN is taken to be 5 cm$^{-1}$ (about $10^{12}$ $s^{-1}$.)  }
\end{figure}



Figure 2 presents a system where the two-photon emission is made dominant over all other decay channels. We consider (for concreteness) the example of a hydrogen atom above a cubic boron nitride (cBN) slab (as in Fig.1), and present the two-photon spectral enhancement. The hydrogen transition we consider is the 5s-4s two-photon transition at 2468 cm$^{-1}$ (4.05 $\mu$m), so that the energies of the emitted polaritons fit in the Reststrahlen (RS) band where the Purcell factor is very high $(> 10^5)$ (Fig. 2a). In Figure 2a, we compute the Purcell spectra for a first-order dipole transition for atom-surface separations of 5, 10, and 25 nm to get an order of magnitude estimate for the rates of the competing dipole transitions at first order. At 5 nm, the fastest competing transition occurs with a lifetime of order 100 ns. At 10 and 25 nm, the order of magnitude of the competing E1 transition is closer to 1 $\mu$s. In Figures 2(b-d), we compute the spectrum of two-photon emission from 5s-4s (due to cBN phonon-polaritons),  the lifetime of the two-photon transition and the $r$-values, the last of which we define as the ratio of the decay rate computed assuming no losses ($\Gamma_R$) to the decay rate computed with losses taken into account ($\Gamma$). It is a measure of the extent to which quenching dominates the decay dynamics insofar as a low r value suggests strong loss-dominated decay. An $r$ value of nearly 1 suggests that losses have little impact on the decay rate. In Figures 2(b-d), we see that the lifetimes of two-photon spontaneous emission for an emitter 5, 10, and 25 nm away from the surface of 10 nm thick cBN are 1.7 ns, 68 ns, and 28 $\mu$s, respectively. Their $r$-values are 0.51, 1.02, and 1.12, respectively. Thus, two-photon spontaneous emission into phonon-polaritons can be nearly two orders of magnitude faster than single-photon dipole transitions. This is in sharp contrast to the situation in free-space, where two-photon spontaneous emission is roughly 8-10 orders of magnitude slower. We thus conclude that using phonon-polaritons, it is possible to create a source of a pair of entangled polaritons with very high efficiency. We now move on to describing the spectral properties of this new quantum light source, both in frequency and angle.
\begin{figure}[t]
\centering
\includegraphics[width=190mm]{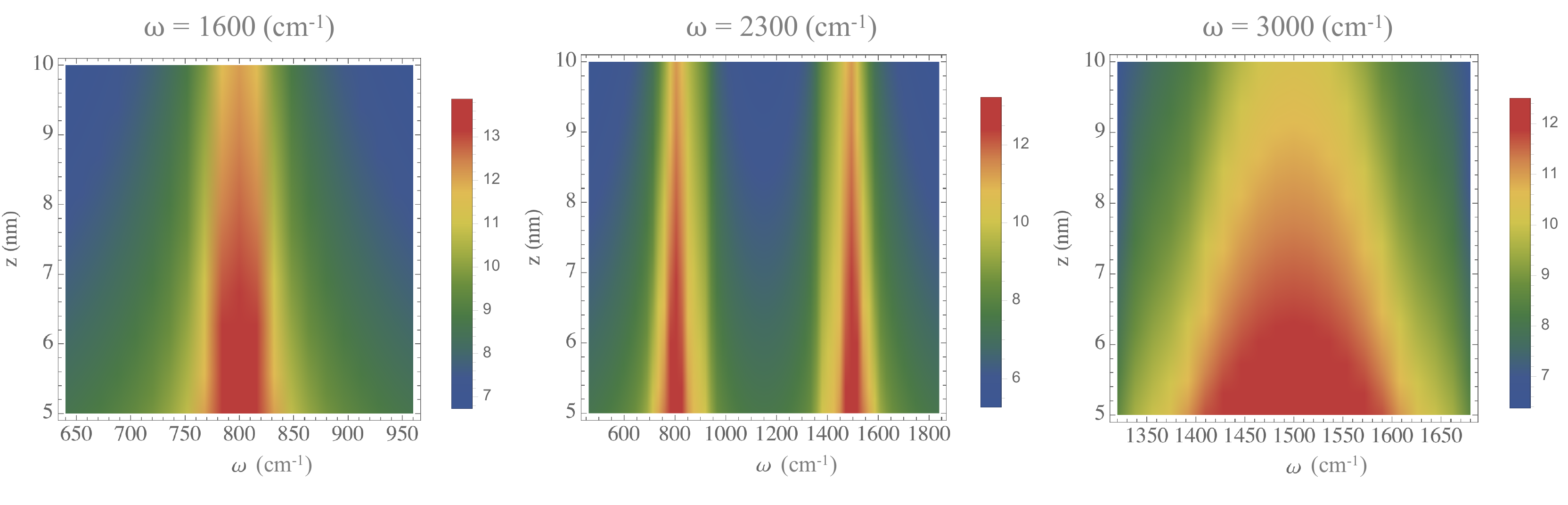}
\caption{\textbf{Hyperbolicity and Two Photon Emitters in Multiple Bands.} Two-polariton spectral enhancement defined as in Equation (1) for a spherical emitter as a function of transition frequency (1600 (left), 2300 (center), and 3000 (right) cm$^{-1}$), plotted on log$_{10}$ scale. The spectral enhancement is plotted with respect to emission frequency and atom-surface separation between 5 and 10 nm. Hyperbolicity allows for enhancement over a large range of frequencies compared to isotropic systems. Moreover, distance can be used to tune the width of the spectrum.}
\end{figure}

In the Figure 3, we also considered the spectral enhancement (plotted on log$_{10}$ scale) defined above in hBN as a function of the transition frequency of the emitter ($\omega_0 = 1600, 2300, 3000$ cm$^{-1}$), emission frequencies, and emitter-surface separation ($z_0 = 5-10$ nm). We chose a different material than that of Figure 2 to show that two-photon spectral enhancements similar to those in thin cBN are achievable in other materials (the spectral enhancement is of the same order of magnitude as that in cBN (about $10^{12}$)) and also to show a large number of frequency bands where a two-photon emitter can be created. What we found is that hBN offers very high spectral enhancement in three different frequency bands, as opposed to one in isotropic polar dielectrics, allowing for compatibility with many more electronic systems. The reason for this is intercombination: a two-photon emission through near-field polaritons can occur via two photons in the lower RS band, two photons in the upper RS band, or one in the upper RS band and one in the lower RS band. A hypothetical hyperbolic material having three separate RS bands would offer six frequency ranges for two-photon emission. We also found that increasing the emitter separation causes the emission spectrum not only to be weaker, but also narrower. This allows one to tune not only the emission rate but also the emission spectrum with atom-surface separation for emitters whose location can potentially be precisely controlled. 

To perform the calculations leading to the results in Figure 2 and Figure 3, we make use of a general result (that we derive in the SM) that for an s $ \rightarrow $ s transition, the spectral enhancement factor, defined as the ratio of the SPhP emission spectrum, $\frac{d\Gamma}{d\omega}$, to the free-space emission spectrum $\frac{d\Gamma_0}{d\omega}$ is given by:
\begin{equation}
\text{Spectral Enhancement }= \frac{\frac{d\Gamma}{d\omega}}{\frac{d\Gamma_0}{d\omega}} = \frac{1}{2}F_p(\omega)F_p(\omega_0-\omega),
\end{equation}
where $F_p(\omega)$ is the Purcell factor for a dipole perpendicular to the surface and is related to the imaginary part of the reflectivity of the air-polar dielectric-substrate system by $F_p(\omega) = \frac{3}{2k_{vac}^3}\int dq ~q^2e^{-2qz_0}\text{Im }r_p(q,\omega)$, $r_p$ is the p-polarized reflectivity of the air-slab-substrate system, and $k_{vac}$ is the free-space photon wavevector, $\frac{\omega}{c}$~\cite{kumar2015tunable}.

Finally, we consider the angular spectrum of emitted photons. In the Supplementary Materials, we derive the general result that the angle and frequency spectrum of two-photon emission, $S(\omega,\theta,\theta')$  is proportional to:
\begin{equation}
S(\omega,\theta,\theta') \sim \Big|\sum\limits_{ij} \hat{e}_i^*(\theta)\hat{e}_j^*(\theta')T_{ij}(\omega) \Big|^2,
\end{equation}
where $T_{ij}(\omega) = T_{ji}(\omega_0-\omega) \sim \sum_n \left(\frac{d_i^{gn}d_j^{ne}}{E_e-E_n-\hbar\omega} + \frac{d_j^{gn}d_i^{ne}}{E_e-E_n-\hbar(\omega_0-\omega)} \right)$,
where $d^{ab}$ denotes a dipole matrix element between states $a$ and $b$, $n$ denotes an intermediate atomic state, $g$ denotes the ground state, $e$ denotes the excited state, and $E_i$ is the energy of the $i$th state. The $\hat{e}_i(\theta)$ are the phonon-polariton polarizations in the vicinity of the emitter, given by $\frac{1}{\sqrt{2}}(\cos\theta,\sin\theta,i)$

\begin{figure}[hh]
\includegraphics[width=160mm]{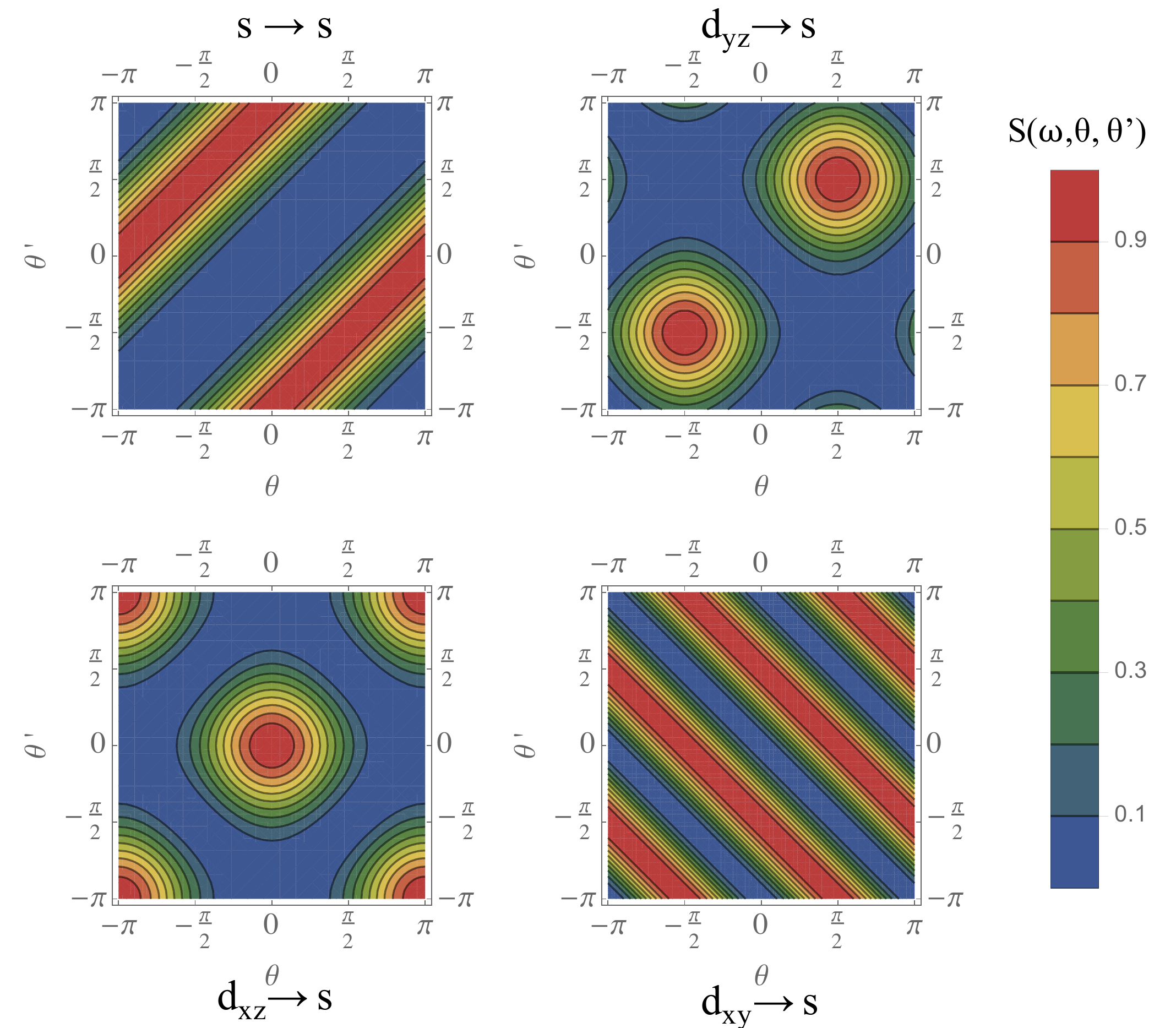}
\caption{\textbf{Using the Shape of the Wavefunction to Control the Angle Spectrum of Emitted Entangled Pairs:} Plots of the angular spectrum $S(\omega_0/2,\theta,\theta')$ of two-photon emission as a function of the initial state of the electron for initial states $s, d_{xy},d_{xz},d_{yz}$.}
\end{figure}

The angular dependence of the spectrum in Equation (2) will lead to very different angular spectra for different transitions. In Table 1, we show the angular spectrum as a function of different transitions (at $\omega = \omega_0/2$). Strictly speaking, the angular spectrum is frequency dependent. However, due to the narrowness of the RS band(s), this can be neglected, and thus we only consider the spectrum at half the transition frequency. Remarkably, just by changing the initial state of the system, one can change whether the entangled pairs are preferentially emitted in the same direction (as in Figure 4(top right,bottom left)) or in opposite directions (as in Figure 4(top left)). There are a number of ways to preferentially populate a particular initial state. One is by exciting the atoms with light of a fixed polarization. Another, appropriate in systems with less extreme degeneracy than hydrogen, is to simply excite the atoms with the appropriate frequency.
\begin{table*}[t]
\centering
\begin{tabular}{|c|c|}
\hline
Transition & Angular Spectrum   \\
\hline
$s \rightarrow s$ & $\sin^4\left(\frac{\theta-\theta'}{2} \right)$ \\
$d_{xy} \rightarrow s$ & $\sin^2(\theta+\theta')$ \\
$d_{xz} \rightarrow s$ & $(\cos\theta+\cos\theta')^2$ \\
$d_{yz} \rightarrow s$ & $(\sin\theta+\sin\theta')^2$\\
\hline
\end{tabular}
\caption{A summary of the dependence of the angular spectrum of two-photon radiation as a function of initial and final electronic states for a few selected initial electronic states.}
\end{table*}

\section{Discussion}

Although we have considered the hydrogen atom in our calculations, this was just for concreteness and the physical mechanism for efficient two-photon emitters can readily be extended to many atomic and molecular systems. Here, we propose some interesting emitter platforms for testing the predictions of our theory. For atoms, there are many which have a level structure conducive to the situation described throughout this paper. For example, in the Lithium atom there is a $d \rightarrow s$ transition at 1611 cm$^{-1}$, which can potentially occur via emission of a pair of SPhPs. All other competing transitions fall well outside of the Reststrahlen bands of hBN, just as in the example we discussed in Figure 2.  More generally, as all atoms have electronic transitions in the mid-IR, our results should apply to a large portion of the periodic table. It may also be possible that vibrational transitions may be used to observe these effects, although one obstacle towards realizing these effects with molecular vibrations is the generally low multipole moments associated with vibrational transitions. Another exciting possibility is using emitters whose level structure is designable, such as quantum dots or wells. There, one can envision designing a two-photon transition to fall in the Reststrahlen bands while designing all other one-photon transitions to fall out of the band. Yet another advantage of these artificial atomic systems is their relatively large sizes, allowing for impedance matching to be realized which much less requisite polariton confinement.

In this work, we focused our attention on quantum optics at mid-IR/THz frequencies; a relatively undeveloped field but nonetheless one which would be rather exciting to achieve. Moving beyond the mid-IR to the near-IR and eventually visible, it may be possible to realize the effects we describe here using 'shaped polaritonic media' (such as nano-resonators of graphene or plasmonic crystals of graphene and other 2D plasmonic materials) supporting (respectively) narrow spectral response or photonic bandgaps. We note that in all of these examples, the emitted quanta are confined modes and do not escape into the far-field unless out-coupled. Out-coupling efficiencies of a few percent were demonstrated in highly confined graphene plasmons, and even higher efficiencies should be demonstrable through optimization. We believe that the results presented in this work may have direct implications for spectroscopy (to infer electronic transitions which cannot be determined with conventional photons), sensors based on forbidden transitions, quantum radiation sources (on-demand generation of single photons and entangled pairs of photons), new platforms for quantum non-linear optics, the possibility of realizing nonlinearities at the single photon level, the ability to turn narrow-band emitters into broadband emitters, the ability to turn narrow-band absorbers into broadband absorbers, and most generally, the ability to completely reshape the ostensibly fixed optical properties of materials.


\section{Methods}
To capture the effect of the optical environment on two-photon emission, fully quantum calculations are necessary. The results presented in this paper take losses into account rigorously through the formalism of macroscopic QED which we use to describe surface phonon polaritons. The two-photon process can be captured by an interaction Hamiltonian ~\cite{tame2013quantum} $H_{int} = -\mathbf{d}\cdot\mathbf{E}$, where the electric field operator, $\mathbf{E}$, in the presence of losses is given by~\cite{knoll2000qed,scheel2008macroscopic}:
\begin{equation}
E_i(\mathbf{r}) = i\sqrt{\frac{\hbar}{\pi\epsilon_0}}\int d\mathbf{r'} \int d\omega' \frac{\omega^{'2}}{c^2} \sum\limits_{k'} \sqrt{\text{Im }\epsilon(\mathbf{r'},\omega')} \left(G_{ik'}(\mathbf{r},\mathbf{r'},\omega')\hat{f}_{k'}(\mathbf{r'},\omega')  - \text{H.c} \right),
\end{equation}
with $G_{ik'}(\mathbf{r},\mathbf{r'},\omega')$ being the dyadic Green function of the Maxwell equations for the material system presented on the right side of Figure 1. The $f^{(\dagger)}_k(\mathbf{r},\omega)$ operators are ladder operators for the fundamental bosonic excitations of an absorbing medium, which can be thought to be dipoles with definite position ($\mathbf{r}$), frequency ($\omega$), and polarization ($k$). They satisfy bosonic commutation relations: $[f_j(\mathbf{r},\omega),f^{\dagger}_k(\mathbf{r}',\omega')] = \delta_{jk}\delta(\mathbf{r}-\mathbf{r}')\delta(\omega-\omega')$.  For the low losses characteristic of phonon-polaritons, decay rates can be obtained with reasonable accuracy by neglecting the losses and writing the field operators in the form of an expansion over plane-wave phonon-polariton modes. We discuss this effective mode expansion in more detail in the Supplementary Materials.  The advantage of the Green function expression is generality and the ability to characterize the impact of losses on decay rates. The advantage of the lossless formalism is that it emphasizes the fact that the emission is into modes, and from this, we can easily extract information such as the angular spectrum of emitted phonon-polaritons, as we do later in the text. We use both formalisms throughout this work.

\section{Acknowledgements}

This work was partly supported by the Army Research Office through the Institute for Soldier
Nanotechnologies under contract no W911NF-13-D-0001. M.S. (analysis and reading of
the manuscript) was supported by S3TEC, an Energy Frontier Research Center funded by the US
Department of Energy under grant no. DE-SC0001299. I.K. was supported in part by Marie Curie grant no. 328853-MC-BSiCS. N.R. was supported by a Department of Energy fellowship no.  DE-FG02-97ER25308. We also thank J. D. Joannopoulos, E. Ippen, J.J. Lopez, and B. Zhen for fruitful discussions.

\section{Supplementary Materials for: A near-unity efficiency source of entangled guided waves}

In this appendix, we derive some results regarding the coupling of atomic emitters to the phonon-polaritions characteristic of polar crystals such as hBN and SiC. In particular, we consider atom-field interactions governed by the non-relativistic Pauli-Schrodinger Hamiltonian $H$:
\begin{align}
&H = H_a + H_{em} + H_{int} \nonumber \\
&H_a = \left(\sum_i \frac{\mathbf{p}_i^2}{2m_e} - \frac{e^2}{4\pi\epsilon_0r_i}\right) \nonumber \\
&H_{em} = \sum_k \int d\mathbf{r} \int d\omega ~ \hbar\omega\left(f_k^{\dagger}(\mathbf{r},\omega)f_k(\mathbf{r},\omega) + \frac{1}{2}\right) \nonumber \\
&H_{int} = \sum_i \frac{e}{2m} (\bold{p}_i\cdot\bold{A}(\mathbf{r}_i)+\bold{A}(\mathbf{r}_i)\cdot\bold{p}_i ) + \frac{e^2}{2m}\bold{A}^2(\mathbf{r}_i),
\end{align}
$H_a$ is the atomic Hamiltonian, $H_{int}$ is the atom-field interaction and  $\bold{A}$ is the vector potential operator. The minimal-coupling interaction Hamiltonian presented above is related to the more well-known dipole interaction Hamiltonian: $-\mathbf{d}\cdot\mathbf{E}$ + self-energy, by a unitary transformation in the long-wavelength (dipole) approximation~\cite{cohen1992atom}. 

The approach that we take accounts for losses via the formalism of macroscopic QED. The primary physical difference between QED without losses and QED with losses lies in the elementary excitations. In the lossless formalism, the excitations can be seen as quanta of electromagnetic modes. In the lossy formalism, the excitations cannot be seen as quanta of electromagnetic modes because the modes are no longer complete. Rather, the elementary excitations are point dipoles which are induced in the material. These excitations are characterized by position, frequency, and orientation. 

The vector potential operator in the framework of macroscopic QED is given by: \begin{equation}
A_i(\mathbf{r})= \sqrt{\frac{\hbar}{\pi\epsilon_0}}\int d\omega'~ \sum\limits_j \frac{\omega'}{c^2} \int d\mathbf{r'} \sqrt{\text{Im }\epsilon(\mathbf{r'},\omega')} G_{ij}(\mathbf{r}, \mathbf{r'} ; \omega')\hat{f}_{j}(\mathbf{r'},\omega') + \text{H.c.},
\end{equation}
where $G_{ij}$ is the dyadic Green function of the Maxwell equations, satisfying $\nabla \times \nabla \times \mathbf{G}_i - \epsilon(\mathbf{r},\omega)\frac{\omega^2}{c^2}\mathbf{G}_i = \delta(\mathbf{r}-\mathbf{r'})\hat{e}_i$.  The operator $\hat{f}_j^{(\dagger)}(\mathbf{r},\omega)$  annihilates (creates) a lossy excitation of frequency $\omega$, at position $\mathbf{r}$, and in direction $j$. It satisfies bosonic commutation relations, namely: 
$\left[ \hat{f}_i(\mathbf{r},\omega), \hat{f}_j^{\dagger}(\mathbf{r'},\omega') \right] = \delta_{ij}\delta(\omega-\omega')\delta(\mathbf{r}-\mathbf{r'})$. When applying the Fermi Golden Rule, the initial state is $|e,0\rangle$, while the final states are of the form $|g,\mathbf{x}_1\omega_1 k_1,...\mathbf{x}_N\omega_N k_N\rangle$ \cite{knoll2000qed}, where $g$ represents a ground atomic state, $e$ represents an excited atomic state, and $|\mathbf{x}\omega k\rangle \equiv \hat{f}_k^{\dagger}(\mathbf{x},\omega)|0\rangle$ represents an excitation of the electromagnetic field.


\section{Optical Response of Phonon-Polariton Materials}

As prescribed by Equation (2), we must compute the Green function of a phonon-polariton supporting system.  The simplest computation involves invoking the excellent approximation that the wavevector of the emitted phonon polaritons  is much larger than the photon wavelength $\frac{\omega}{c}$. This is the electrostatic limit. 

We can compute the Green function for an both an anisotropic and an isotropic polar crystal with the same methods. As is well known for a 2D-translationally-invariant system, the Green function is most easily determined by writing it in Fourier space (decomposed into parallel wavevectors) and solving the Maxwell equations for each Fourier component. This Fourier integral  is computed for the p-polarized polaritons; the s-polarized modes give a very weak contribution in the electrostatic limit. In the region above the dielectric slab, the Green function is known once one finds the p-polarized reflectivity of the system, $r_p$. In particular the p-polarized green function takes the form \cite{scheel2008macroscopic}:
\begin{equation}
G_{ij}(\mathbf{r},\mathbf{r'},\omega) = \frac{i}{2}\frac{1}{(2\pi)^2} \int d\mathbf{q}~C^p_{ij}e^{i\mathbf{q}\cdot\boldsymbol{\rho}+ik_{\perp}z}e^{-i\mathbf{q}\cdot\boldsymbol{\rho}'+ik_{\perp}z'}.
\end{equation}
In the electrostatic limit relevant to our calculations, the p-polarized reflectivity takes the form \cite{Rivera2016}:
$$C^p_{ij}=-2ic^2\frac{r_pq}{\omega^2}\hat{\epsilon}_i(\mathbf{q})\hat{\epsilon}_j(\mathbf{q})^*,$$
where the polarization vectors are defined by: $\hat{\boldsymbol{\epsilon}}(\mathbf{q}) \equiv \frac{\hat{\mathbf{q}}+i\hat{\mathbf{z}}}{\sqrt{2}}$. One finds that for an isotropic slab of polar dielectric of thickness $d$ and permittivity $\epsilon$ with a superstrate of air and a non-dispersive substrate of dielectric constant $\epsilon_s$ that:
\begin{equation}
r_p = \frac{e^{2 qd}(\epsilon -1) (\epsilon_s+\epsilon )+(\epsilon +1) (\epsilon_s-\epsilon )}{e^{2 qd}(\epsilon +1) 
   (\epsilon_s+\epsilon )+(\epsilon -1) (\epsilon_s-\epsilon )}.
\end{equation}
Similarly one finds that for an anisotropic slab of polar dielectric (like hBN) of thickness $d$ and permittivity $\text{diag}(\epsilon_{\perp},\epsilon_{\perp},\epsilon_{||})$ with a superstrate of air and a non-dispersive substrate of dielectric constant $\epsilon_s$ that:
\begin{equation}
r_p = \frac{\left(\sqrt{r} \epsilon_{||} +i\right) \left(\epsilon_s+i \sqrt{r} \epsilon_{||} \right) e^{2 i q \sqrt{r} d}+\left(\sqrt{r} \epsilon_{||}
   -i\right) \left(\epsilon_s-i \sqrt{r} \epsilon_{||} \right)}{\left(\sqrt{r} \epsilon_{||} -i\right) \left(\epsilon_s+i \sqrt{r} \epsilon_{||}
   \right) e^{2 i q \sqrt{r} d}+\left(\sqrt{r} \epsilon_{||} +i\right) \left(\epsilon_s-i \sqrt{r} \epsilon_{||} \right)}
\end{equation}
where $r$ is the absolute value of the anisotropy ratio, defined by $r= \Big|\frac{\epsilon_{\perp}}{\epsilon_{||}}\Big|$. The location of the poles of the imaginary part of the reflectivity in $(\omega,q)$ space gives the dispersion relation $\omega(q)$. When losses are present, Im $r_p$ is centered around the dispersion relation.

\section{Macroscopic QED at Higher Order in Perturbation Theory: Emission of Two Polaritons}
In this section, we derive the frequency spectra of spontaneous emission of two excitations of the lossy electromagnetic field.  These results incorporate losses and thus elucidate the contribution of quenching and polariton launching to the decay of an excited emitter. The derivation proceeds by application of the Fermi Golden rule at second order in perturbation theory as applied to transitions between an initial state $|e,0\rangle$ and a continuum of final states with two excitations of the lossy electromagnetic field, i.e., $|g,\mathbf{x}\omega k,\mathbf{x}'\omega'k'\rangle.$ Fermi's Golden Rule for this decay reads:
\begin{equation}
\Gamma = \frac{2\pi}{\hbar^2}\frac{1}{2}\int d\mathbf{r}d\mathbf{r'}\int d\omega d\omega' \sum\limits_{k,k'} \Big|\sum_{i_1} \frac{\langle g, \mathbf{r}\omega k,\mathbf{r'}\omega' k'|V|i_1 \rangle \langle i_1|V|e,0 \rangle}{E_e - E_{i_1} + i0^+} \Big|^2\delta(\omega_0 - \omega - \omega'),
\end{equation}
where $|i_1\rangle$ are intermediate states containing both the atom and field degrees of freedom. The sum is understood to be a sum over discrete degrees of freedom and an integral over continuous ones. The factor of $1/2$ comes from the fact that when we sum over all $\{\mathbf{r}\omega k,\mathbf{r'}\omega' k\}$ pairs, each pair of excitations appears twice. To proceed, we express the field operators in terms of Green functions and use three facts:

1. \begin{align}
V_{i_j,i_{j-1}} &= \langle n_{j},\mathbf{x}_j\omega_j k_j,\mathbf{x}_{j-1}\omega_{j-1}k_{j-1},..,.\mathbf{x}_{1}\omega_{1}k_{1}  |d_iE_i|n_{j-1},\mathbf{x}_{j-1}\omega_{j-1}k_{j-1},...,\mathbf{x}_{1}\omega_{1}k_{1} \rangle \nonumber \\ &= i\sqrt{\frac{\hbar}{\pi\epsilon_0}}d^{n_j,n_{j-1}}_{i_j}\frac{\omega_j^2}{c^2}\sqrt{\text{Im }\epsilon(\mathbf{x}_j,\omega_j})G_{i_j,k_j}^*(\mathbf{r}_0,\mathbf{x}_j,\omega_j),\nonumber
\end{align}

2. $$
\frac{\omega^2}{c^2} \int d\mathbf{r}~ \text{Im }\epsilon(\mathbf{r},\omega) (GG^{\dagger})(\mathbf{r}_0,\mathbf{r},\omega) = \text{Im }G(\mathbf{r}_0,\mathbf{r}_0,\omega), \text{and}
$$

3. For a photonic medium which is translationally invariant in-plane (as is the system we consider throughout the text):
$$
\text{Im }G_{ij}(\mathbf{r}_0,\mathbf{r}_0,\omega) = \frac{\omega}{6\pi c}F_p(\mathbf{r}_0,\mathbf{r}_0,\omega) D_{ij},
$$
where $D = \text{diag}(1/2,1/2,1)$ and $F_p$ is the Purcell factor for one-photon emission for the z-polarized dipole (frequency $\omega$ and position $\mathbf{r}_0$) near this material \cite{koenderink2010use}. As a reminder, this Purcell factor is equal to $\frac{3c^3}{2\omega^3}\int dq~q^2e^{-2qz_0}\text{Im }r_p(q,\omega)$.
%
%

Defining (note that this definition differs from that in the main text by a factor of the squared electron charge) $$ T_{ij}(\omega) = \sum\limits_n \frac{x_j^{gn}x_i^{ne}}{\omega_e-\omega_n-\omega} + \frac{x_i^{gn}x_j^{ne}}{\omega_e-\omega_n-(\omega_0-\omega)} = T_{ji}(\omega_0-\omega), $$ we see that the second-order emission spectrum becomes:
\begin{align}
\frac{d\Gamma}{d\omega}=\frac{4\alpha^2}{9\pi c^4} ~&\omega^3(\omega_0-\omega)^3F_p(\omega)F_p(\omega_0-\omega) \sum\limits_{i,j,r,s}D_{ri}D_{sj}T_{ij}(\omega)T_{rs}(\omega)^*,
\end{align}
Because $D$ is diagonal, this is simply:
\begin{equation}
\frac{d\Gamma}{d\omega} = \frac{4\alpha^2}{9\pi c^4} \omega^3(\omega_0-\omega)^3F_p(\omega)F_p(\omega_0-\omega) \sum\limits_{ij} D_{ii} D_{jj} |T_{ij}|^2.
\end{equation}
We focus on the case in which the transition is between two $s$ states. In that case, only the diagonal terms of $T_{ij}$ are relevant meaning that the above sum over $i,j$ becomes $\frac{3}{2}T_{zz}$, making the differential emission rate for two lossy excitations:
\begin{equation}
\frac{d\Gamma}{d\omega} = \frac{2}{3\pi c^4}  \alpha^2 \omega^3(\omega_0-\omega)^3F_p(\omega)F_p(\omega_0-\omega)|T_{zz}|^2
\end{equation}
Using the fact that the free-space differential decay rate is given by \cite{breit1940metastability}:
$$
\frac{d\Gamma_0}{d\omega} = \frac{4}{3\pi c^4}\alpha^2\omega^3(\omega_0-\omega)^3|T_{zz}|^2,
$$
it follows that the spectral enhancement (defined in the main text) is:
\begin{equation}
\frac{\frac{d\Gamma}{d\omega}}{\frac{d\Gamma_0}{d\omega}} = \frac{1}{2}F_p(\omega)F_p(\omega_0-\omega).
\end{equation}
By evaluating the $T_{ij}$ tensors, we can easily go from the spectral enhancement factors to the two-photon Purcell factor by evaluating the sum over states, as we do in order to get the rates claimed in the main text. Moreover, the actual decay rate can easily be estimated in the case where the bandwidth of the polaritons is narrow (as is the case in what we consider) as:
\begin{equation}
\Gamma \approx \frac{\alpha^2}{96\pi c^4}\omega_0^6\left|T_{zz}\left(\frac{\omega_0}{2}\right)\right|^2\int d\omega ~F_p(\omega)F_p(\omega_0-\omega),
\end{equation}
in the case where the emission spectrum is sharply centered at $\frac{\omega_0}{2}$ as a result of the sharpness of the spectral enhancement.


\section{Lossless Limit: Effective Mode Expansion}

This section develops a mode expansion formalism, comparing it to the Green function formalism used so far, and discusses their relative strengths and weaknesses. Although all of the decay rates that we compute can be computed through the Green function formalism presented in the previous section, it is difficult to extract information such as the angular spectrum of emitted polaritons from this formalism (which we define as the angular spectrum in the lossless limit). The reason for this difficulty is that concepts like the angular spectrum are clearly most naturally computed when one assumes that the excitations coupled to are modes labeled by their direction of propagation. Therefore, to compute the angular spectrum of emitted radiation, we use a formalism different from the one used in the previous sections. We will write down field operators appropriate to the lossless situation and compute the spectrum of two-polariton emission by using Fermi's Golden Rule with these field operators.
 
It is known that in lossless and non-dispersive dielectrics, the vector potential can be expressed in the form of a mode expansion:
$$
\mathbf{A} = \sum\limits_{n}\sqrt{\frac{\hbar}{2\epsilon_0\omega_n}} \left(\mathbf{F}_na_n + hc \right),
$$
where the $\mathbf{F}_n$ are the orthonormal modes of the Maxwell equations, normalized suitably. In \cite{lin2016tailoring}, it was rigorously shown by taking the Green function formalism in the lossless limit that a mode expansion for the field operators in terms of eigenmodes (of the form above) can be derived for polaritons. In this effective mode expansion, the field modes are normalized such that:
$$
\frac{\epsilon_0}{2\omega}\int d\mathbf{r}~ \mathbf{F}^*(\mathbf{r})\cdot\frac{d(\epsilon_r\omega^2)}{d\omega}\cdot\mathbf{F}(\mathbf{r}) = \frac{\hbar\omega}{2}.
$$
We take as the normalization or quantization volume one which is infinite in the z-direction and has area 1 $m^2$ in the in-plane direction. In the electrostatic limit $\frac{qc}{\omega}\gg 1$, the fields in the vicinity of a well-localized emitter above a polar dielectric are of the form \begin{equation}\mathbf{F} \sim e^{i\mathbf{q}\cdot\boldsymbol{\rho}-qz}\hat{e}(\hat{\mathbf{q}}),\end{equation} where $\hat{e}(\hat{\mathbf{q}})\equiv \frac{\hat{\mathbf{q}}+i\hat{\mathbf{z}}}{\sqrt{2}}$. $\hat{\mathbf{q}}$ can be expressed as $\cos\theta\hat{\mathbf{x}}+\sin\theta\hat{\mathbf{y}}$. We use this fact to compute the angular spectrum of pairs of emitted phonon-polaritons in what follows.

We conclude this part of the discussion by noting that effective mode expansion was shown by proving that the denominator of the Fourier transformed Green function in the lossless limit is proportional to the energy term in the previous equation. Although we derived this result on very general grounds (see \cite{lin2016tailoring}), we explicitly show the equivalence here as a consistency check on our calculations. In Figure S1, we compare predictions of the mode-expansion formalism and the Green function formalism taken in the zero loss limit. The particular prediction we address is the Purcell factor of a z-polarized dipole some distance away from hexagonal boron nitride ((a) and (b)) or cubic boron nitride (c). As can be seen, aside from small numerical integration error, these predictions match extremely well. In Figure S2, we consider the same Purcell factor but now we compare the lossless value of the Purcell factor to the Purcell factor when realistic losses are incorporated into the Lorentz permittivities of hexagonal boron nitride and cubic boron nitride ($\gamma = 5$ cm$^{-1}$ is taken in all three cases). We can see that these predictions agree reasonably well.

\begin{figure}[hh]
\centering
\includegraphics[width=160mm]{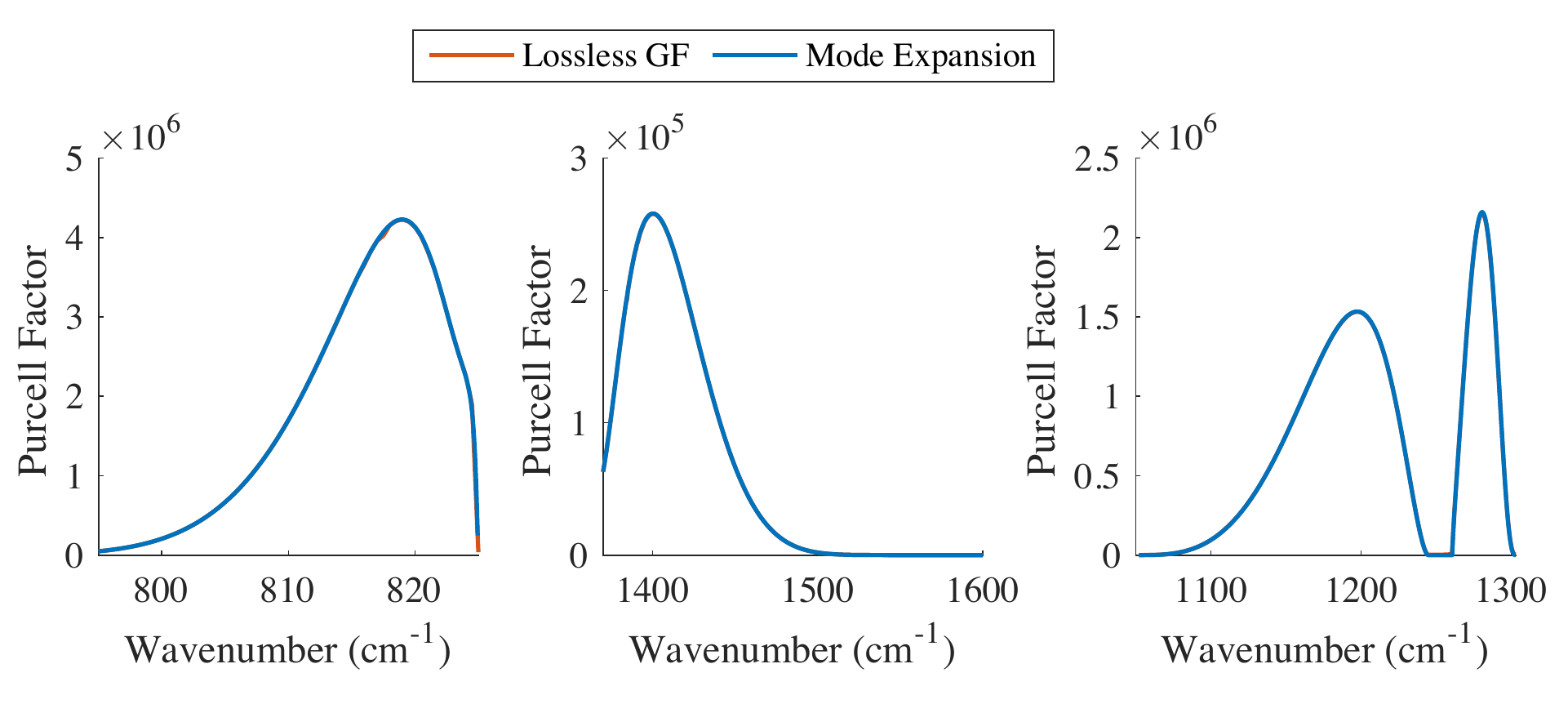}
\caption{\textbf{Comparison of Lossless Green Function Formalism and Mode Expansion} Purcell Factors calculated for (a) the lower Reststrahlen band of hBN for an hBN thickness of 2 nm and emitter-surface separation of 10 nm (b) the upper Reststrahlen band of hBN for an hBN thickness of 2 nm and emitter-surface separation of 10 nm (c) the Reststrahlen band of cBN for a cBN thickness of 5 nm and emitter-surface separation of 10 nm.}
\end{figure}

\begin{figure}[hh]
\centering
\includegraphics[width=160mm]{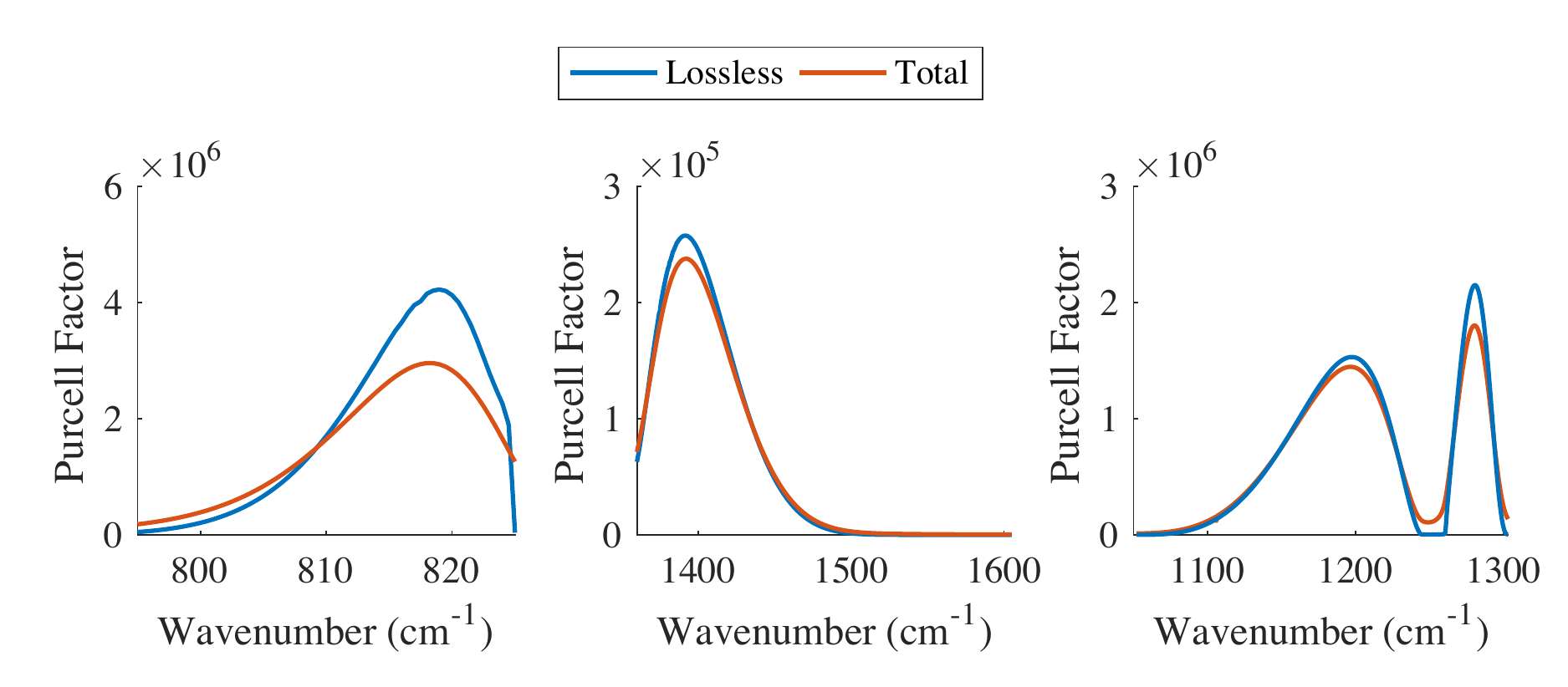}
\caption {\textbf{Comparison of Green Function Formalism with Losses and Mode Expansion}  Purcell Factors calculated for (a) the lower Reststrahlen band of hBN for an hBN thickness of 2 nm and emitter-surface separation of 10 nm (b) the upper Reststrahlen band of hBN for an hBN thickness of 2 nm and emitter-surface separation of 10 nm (c) the Reststrahlen band of cBN for a cBN thickness of 5 nm and emitter-surface separation of 10 nm.}
\end{figure}

\subsection{Angular Spectrum of Emitted Phonon Polaritons}

Now, we focus on computing the angular spectrum of radiation of phonon-polaritons emitted by an excited atomic electron. In other words, we want the quantity $$S(\omega,\theta,\theta') \equiv \frac{d\Gamma}{d\omega d\theta d\theta'}.$$ Because we want to focus only on excitation of propagating polaritons and not  loss-excitations, we extract the pole contribution from the imaginary part of the p-polarized reflectivity. This is equivalent to writing field operators in the lossless limit.  Writing the second-order Fermi Golden Rule for the transition rate between an initial state $|e,0\rangle$ and the continuum of final states $|g,\mathbf{q}\mathbf{q'}\rangle$, we see that:
$$
\frac{d\Gamma}{d\omega d\theta d\theta'} = \frac{1}{16\pi^3\hbar^2} \frac{q(\omega)q(\omega_0-\omega)}{v_g(\omega)v_g(\omega_0-\omega)}\Big|\sum\limits_{i_1} \frac{\langle g,\mathbf{q}\mathbf{q'}|\mathbf{d}\cdot\mathbf{E}|i_1\rangle\langle i_1|\mathbf{d}\cdot\mathbf{E}|e,0\rangle}{E_e-E_{i_1}+i0^+} \Big|^2,
$$
where $v_g$ is the group velocity, $\frac{d\omega}{dq}$. Inserting the definition of the operators, we find that the spectrum is given by:
\begin{equation}
S(\omega,\theta,\theta') = \frac{\alpha^2c^2}{4\pi}\omega(\omega_0-\omega) \frac{q(\omega)q(\omega_0-\omega)}{v_g(\omega)v_g(\omega_0-\omega)}\Big| F_{\mathbf{q}}^{*i}F_{\mathbf{q'}}^{*j}T_{ij} \Big|^2
\end{equation}
where $$ T_{ij}(\omega) = \sum\limits_n \frac{x_j^{gn}x_i^{ne}}{\omega_i-\omega_n-\omega} + \frac{x_i^{gn}x_j^{ne}}{\omega_i-\omega_n-(\omega_0-\omega)} = T_{ji}(\omega_0-\omega). $$

We now use this to extract the form of the angular spectrum of entangled photons as a function of the electronic orbitals participating in the transition. To give the reader a sense of how much control one may have over the angular spectrum of emitted photon pairs, we consider four cases. In all four, the final states are $s$ states. But the initial states will be taken to be $s, d_{xy}, d_{yz},$ and $d_{xz}$ states. 
\subsubsection{s $\rightarrow$ s}
In the case where the initial state is an $s$ state, we have that $T_{ij}=0$ if $i \neq j$. This is because of the dipole approximation, which fixes the intermediate state to be a p state. Therefore, if $i\neq j$, then $T_{ij}$ has a sum of terms like $\langle s | x_i |p_k\rangle\langle p_k | x_j|s\rangle$, where $p_k = p_x, p_y, p_z$. Each of these terms individually vanishes, and so the entire tensor vanishes. Moreover $T_{xx}=T_{yy}=T_{zz}\equiv T$ because $\langle p_x|x|s\rangle = \langle p_y|y|s\rangle = \langle p_z|z|s\rangle$: Therefore:
\begin{equation}
S(\omega,\theta,\theta') = |T|^2\left(\cos\theta\cos\theta' + \sin\theta\sin\theta'  -1 \right)^2 = 4|T|^2\sin^4\left(\frac{\theta-\theta'}{2} \right). ~(s \rightarrow s)
\end{equation}
\subsubsection{$d_{xy}$ $\rightarrow$ s}
In the case where the initial state is $d_{xy}$, the only contributing terms are $T_{xy}$ and $T_{yx}$. The argument for this statement  makes use of the fact that the $d_{xy}$ has an angular dependence that can be written in Cartesian coordinates as $xy$. We start proving this claim by examing the $T_{zi}$ components. If one of the indices is $z$, then it will either be the case that the intermediate state must be a $p_z$ state (to have overlap with the $s$ state), or that there will be a matrix element of the form $\langle p_i|z|d_{xy}\rangle$. The first case gives zero because $d_{xy}$ has no transition dipole moment with $z$.  The second case also gives zero because $d_{xy}$ has no z-polarized dipole moment with any $p$ orbital. Thus the $T_{zi}$ components vanish. The $T_{xx}$ and $T_{yy}$ components also vanish because $d_{xy}$ has no (x,y)-polarized dipole moment with $p_{x,y}$. Therefore:
\begin{equation}
S(\omega,\theta,\theta') = \left(T_{xy}(\omega)\cos\theta\sin\theta' + T_{xy}(\omega_0-\omega)\sin\theta\cos\theta' \right)^2. ~(d_{xy} \rightarrow s)
\end{equation}

\subsubsection{$d_{xz} \rightarrow s$ and $d_{yz} \rightarrow s$}

A nearly identical argument to the one above (replace all y's with z's or all x's with z's) yields:
\begin{equation}
S(\omega,\theta,\theta') = \left(T_{xz}(\omega)\cos\theta + T_{xz}(\omega_0-\omega)\cos\theta' \right)^2. ~(d_{xz} \rightarrow s)
\end{equation}
\begin{equation}
S(\omega,\theta,\theta') = \left(T_{yz}(\omega)\sin\theta + T_{yz}(\omega_0-\omega)\sin\theta' \right)^2. ~(d_{yz} \rightarrow s)
\end{equation}

\bibliographystyle{unsrt}
\bibliography{thesisbib}
\end{document}